# Role of the dispersion force in modeling the interfacial properties of molecule-metal interfaces: adsorption of thiophene on copper surfaces

Zhi-Xin Hu,[1,2] Haiping Lan,[1,3] Wei Ji[1,2,*]

[1]Department of Physics, Renmin University of China, Beijing 100872, China

[2]Beijing Key Laboratory of Optoelectronic Functional Materials & Micro-Nano Devices, Renmin University of China, Beijing 100872, China

[3]International Center for Quantum Design of Functional Materials, University of Science and Technology of China, Hefei, Anhui 230026, China

[*]wji@ruc.edu.cn; http://sim.phys.ruc.edu.cn/

**We present density functional theory calculations of the geometry, adsorption energy and electronic structure of thiophene adsorbed on Cu(111), Cu(110) and Cu(100) surfaces. Our calculations employ dispersion corrections and self-consistent van der Waals density functionals (vdW-DFs). In terms of speed and accuracy, we find that the dispersion-energy-corrected Revised Perdue-Burke-Enzerhof (RPBE) functional is the ``best balanced" method for predicting structural and energetic properties, while vdW-DF is also highly accurate if a proper exchange functional is used. Discrepancies between theory and experiment in molecular geometry can be solved by considering x-ray generated core-holes. However, the discrepancy concerning the adsorption site for thiophene/Cu(100) remains unresolved and requires both further experiments and deeper theoretical analysis. For all the interfaces, the PBE functional reveals a covalent bonding picture which the inclusion of dispersive contributions does not change to a vdW one. Our results provide a comprehensive understanding of the role of dispersive forces in modelling molecule-metal interfaces.**

Molecular nanostructures on solid surfaces, especially on metals, have received great attention in recent years [1-5]. Accurate modelling of the properties of organic molecule-metal interfaces is of great importance to build, modulate, and utilize these molecular nanostructures [1-8]. Because some of the constituent molecules have a small or vanishing electrical dipole moment, the dispersion force is a major part for the van der Waals (vdW) force between these molecules and other nanostructures. Dispersion forces are not included in conventional density functional theory (DFT), which poses a challenge to the theoretical description of the interfacial properties of these molecules on solid surfaces. However, the inclusion of dispersion forces has been realized in recent developments of DFT [9-20] and has been applied to a number of organic molecule-metal interfaces [21-27].

Still, the role of the dispersion correction in determining the geometry, adsorption energy and bonding picture of an interface remains poorly understood. Dispersion forces have been modelled in Perylene-3,4,9,10-tetracarboxylicacid dianhydride (PTCDA)/Ag(111) [21], thiophene/Cu(110) [22], pyrazine/Cu(110) [23], azobenzene/Ag(111) [24], and $C_{60}$/Au(111) [25] using semi-empirical dispersion corrections (DFT-D) [10-13,24,25], an ab-initio $C_6$ parameter with the consideration of many-body screening effects [26], and van der Waals density functionals (vdW-DF) [17,18], both by a post-Generalized Gradient Approximation (GGA) method [21-24,27] and later by a self-consistent method [24] with a strongly reduced computational cost [28]. To analyze geometries, experimental results from X-ray standing waves (XSW) for PTCDA/Ag(111) [29] and azobenzene/Ag(111) [30] are available for comparison with theory. Neither vdW-DF nor DFT-D reproduces accurately the XSW experiment on PTCDA/Ag(111) [21], but an ionic final state (IFS) approximation [31-33] that considers electron-core-hole interactions does. For azobenzene/Ag(111), out of two DFT-D and four vdW-DF methods, only the Tkatchenko-Scheffler (TS) scheme [13] yields a reasonable Ag-N distance [30]. It is still an open issue whether the XSW measured results are for the ground state [31-34]. If the electronic relaxation after core-hole creation is sufficiently fast, the measured geometry should be identical to the ground-state structure, or if the molecule-substrate bonding is sufficiently strong then the geometry should be very close to that of the ground state. Otherwise, an excited-state method such as IFS should be considered to reproduce the XSW measurements.

An accurate description of adsorption energies has recently been achieved by vdW-DF methods [21,22,24,35]. The choice of the exchange functional together with the vdW correlation functional was found to be of vital importance for an accurate calculation of interaction energies and bond lengths [18-20,35]. These approaches seek the best-match exchange functional to use together with the vdW correlation functional and have generated new exchange functionals including C09 [36], optPBE [19], optB88 [19] and optB86b [20]. The non-local correlation functional can also be optimized, as in

vdW-DF2 [18]. These exchange functionals have usually been tested by a ``standard test" of the S22 set for molecules [19], and by lattice constants, bulk moduli and atomization energies for solids [20]. Although these functionals have also been tested recently for interfaces, including graphene and Ni(111) [35], their performance for organic molecule-metal interfaces remains unknown. In addition, the role of the exchange functional in DFT-D, a much cheaper method, has not yet been appreciated. . The dispersion correction has mostly been employed together with the PBE functional [37], which overestimates adsorption energies quite considerably. In fact different theoretical results remain in conflict over the appropriate bonding picture. Covalent bonding was identified using standard DFT for both pyrazine/Cu(110) [38] and C60/Au(111) [39], and the bonding picture of C60/Au(111) revealed by vdW-DF [27] is also covalent. However, for pyrazine/Cu(110) the bonding was found [23] to be due solely to dispersion forces, contradicting the DFT result.

Resolving such a complex situation calls for a systematic investigation. Because a large amount of experimental data is available for these systems [40-44], in this paper we report DFT calculations for thiophene on Cu(111), Cu(110) and Cu(100) surfaces. We consider the dispersion force at the DFT-D and self-consistent vdW-DF levels, and include the electron-core-hole interaction by the IFS approximation. The geometry of the adsorbed thiophene molecule was relaxed using standard PBE, different DFT-D and vdW-DF methods and standard PBE with the IFS approximation. Using the calculated interface structures, the corresponding adsorption energies were computed using PBE and all DFT-D and vdW-DF methods as were the electronic structures. The local density of states (LDOS) and differential charge density (DCD) were found to support unambiguously a covalent bonding picture for thiophene/Cu, in contrast to a previous suggestion of vdW bonding for thiophene/Cu(110) [22]. The choice of methods (functionals) used in the prediction of structural and energetic properties for these interfaces is summarized in the discussion section. Our results improve the overall understanding of the role of dispersion forces and electron-core-hole interactions in organic molecule-metal interfaces.

## Results

**Structural properties of thiophene/Cu(111).** Our calculated results for the structural relaxations of thiophene adsorbed on three Cu surfaces are summarized in Table 1 and Figure 1. Six configurations of adsorption site and molecular orientation were considered. The top site, where the S atom is located on the top of a Cu atom, shown in Figure 1(a), was found to be the most energetically favored, which is consistent with experiment [40,41]. In experiments, the tilt angle α between the molecular plane of thiophene and the Cu surface varies from 12° to 45° when the coverage increases from 0.03ML to 0.14ML [40]. Because the Cu(111) surface was modeled using a 3×3 supercell, equivalent to a coverage of 0.11 ML, the results should be comparable with the tilting angle α = 26±5° measured at 0.1 ML [40]. The PBE calculation gives 21.6°, within the experimental error bar. Although the PBE Cu-S bond length of 2.68Å is slightly larger than the experimental values of 2.62±0.03Å [40] and 2.50±0.02Å [41], this approach is known to be prone to overestimate bond lengths..

An attractive potential, such as the dispersion correction with semi-empirical pairwise $C_6$-$R_6$ coefficients introduced by Grimme (G06) [11], overcomes the overestimated Cu-S distance. It slightly shortens the Cu-S bond length to 2.51 Å, but drastically reduces the tilting angle to 7.6° [Figure 1(c)]. Two categories of interactions can be inferred from the PBE results, namely Cu-S covalent bonding and the relatively weak Cu-π interaction. The Cu-π interaction cannot adequately resist the attraction from the dispersion correction, leading to a tilting angle significantly smaller than the experiment. The $C_6$-$R_6$ coefficients depend on the atoms involved but not on the functional employed in the calculation. To balance the over-enhanced Cu-π interaction, we employed a more repulsive functional, namely RPBE (RPBE-G06). One flat and one tilted configuration were found to be energetically degenerate (within 3 meV), which implies a competition between the Cu-S and Cu-π interactions. Similar behaviour was also found in the optPBE-vdW and optB88-vdW results. As shown in Figure 1(d) and (e) and Table 1, the flat configuration has a negative tilting angle and a rather large Cu-S bond length, $α_{RPBE-G06-flat}$ = - 1.5° and $d_{RPBE-G06-flat}$ = 3.17Å; in the tilted configuration the bond length, $d_{RPBE-G06-tilted}$ = 2.62Å, is very close to experiment but the angle $α_{RPBE-G06-flat}$ = 13.2° is roughly 10° too small. A similar degeneracy of tilted and flat configurations was previously reported in a DFT investigation of thiophene on Cu(100)[50]. The $C_6$ coefficients in the TS dispersion correction are based on the ground-state electron density [13], which offers a better transferability than G06. However, the replacement of G06 by TS does not cause a qualitative change in the relaxed atomic structures, increasing the Cu-S bond length by only 5-10%.

Although these PBE and DFT-D results are reasonable, they are not satisfactory for predictive purposes. Thus we adopted a more sophisticated method to model the thiophene/Cu(111) interface, namely self-consistent vdW-DF including four combinations of exchange and non-local correlation functionals,. Three of the exchange functionals were used together with the same correlation functional, namely optPBE-vdW, optB88-vdW and optB86b-vdW, while vdW-DF2 is an exception with a non-local correlation functional which was optimized especially for organic molecules. The geometry revealed by the first three functionals is similar to that of RPBE-G06: in the tilted configuration, the Cu-S bond length varies from the shortest value of 2.53Å for optB86b-vdW to the longest value of 2.88Å for optPBE-vdW. The trend of the bond lengths is consistent with the nature of these exchange functionals, in that the most attractive is optB86b and the least attractive optPBE; this result may be read from the exchange enhancement factor $F_x$ as a function of the dimensionless density gradient

$s$ in the range between 1.0 and 2.0 [19,20]. The vdW-DF2 functional performs very well for the S22 set [18] owing to its optimization for molecules, but it fails to reproduce the thiophene/Cu(111) structure, giving rise to a rather large adsorption height, which can also be deduced from $F_x(s)$, and to a negative tilt angle. The angles predicted by the other three methods are also significantly lower than the experimental result (Table 1).

The fact that none of these sets of theoretical bond lengths and tilting angles achieves good agreement with experiment implies that the XSW measurements may reflect the structure of the interface in an effectively charged state. Thus we considered the effects of electron-core-hole interactions by introducing three different effective charges of 0.5$e$, 1.0$e$ and 1.5$e$. This effective charge causes an abrupt change of the adsorption configuration: with an effective charge of 0.5$e$, the molecule moves away from the surface ($d$ = 3.05 Å) and becomes more flat ($a$ = 14.8°). A charge of 1.0$e$ pushes the molecule back to the tilted configuration, which suggests that the screening charges are transferred into a Cu-S bonding state, strengthening the Cu-S covalent interaction. Such an enhanced Cu-S interaction results in an $sp^2$ to $sp^3$ transition of the sulphur hybridization, leading to a more tilted molecule. The bond length $d$ = 2.56 Å and the tilting angle $a$ = 22.0°, shown in Table 1, are both within the error bars of the experiments [40-42].

**Structural properties of thiophene/Cu(110).** The top site is, again, the most favorable adsorption site for thiophene/Cu(110). The Cu-S bond length of 2.39Å (PBE value, see Figure 1(h)) indicates a covalent bond, which differs by only 0.02Å from the previously reported theoretical value of 2.41 Å [22]. The relaxed atomic structure shows that an adjacent C, marked by a red square in Figure 1(g), also bonds covalently to the Cu atom underneath, a result confirmed by the differential charge densities, as elucidated later.

None of the methods more advanced than PBE (listed in Table 1), except vdW-DF2, find significant changes to the Cu-S bond length, which ranges from 2.32 Å to 2.45 Å. This is distinctively different from the case of thiophene/Cu(111), where the tilt angle and bond length appear to be very sensitive to the method. The shorter bond lengths suggest much stronger covalent Cu-S and Cu-C bonds. Thus the same energy correction due to dispersive contributions is minor compared with the bonding energy, resulting in much smaller changes of bond length than on the (111) surface. The IFS estimate of the core-hole interaction, which obviously changes the adsorption geometry of thiophene/Cu(111), shortens the Cu-S bond length only slightly, from 2.40 Å (0.5$e$) to 2.30 Å (1.5$e$). Only the vdW-DF2 method predicts a Cu-S bond length at least 0.3 Å larger than the others, similar to the (111) cases. The tilt angle varies from 3.0° (PBE-G06) to 20.7° (PBE-CH-1$e$) between methods. This wide range of angles can be ascribed to a transition in electronic hybridization from $sp^2$ to $sp^3$ on the carbon adjacent to the S atom, as shown in Figure 1(k) and similar to the case of CuPcF$_{16}$/Ag(111) [31].

No structural information is available from experiment for this interface. The arrangement of atoms on Cu(110) shares with Cu(100) the feature that their $[1\bar{1}0]$ directions are identical, and so measured values for thiophene/Cu(100) may provide a meaningful comparison with the calculated results for Cu(110). The experimental bond length in thiophene/Cu(100) is 2.42±0.02Å [44], which falls in the range from 2.32 Å to 2.45 Å of our theoretical values (other than vdW-DF2), implying good agreement on the Cu-S bond length.

**Structural properties of thiophene/Cu(100).** Early theoretical studies of this interface predicted the Top site to be the most favorable for adsorption [50]. Experiments, however, suggest that the Bridge site is preferred. In our calculations the Top site (Figure 1(l) left) is at least 30 meV more stable than the Bridge site (Figure 1(l) right) for all methods. With the exception of vdW-DF2, the bond length of the Top site configuration is insensitive to the different corrections considered (Table 1), similar to thiophene/Cu(110). The IFS method returns similar structures for all effective charges. At the Bridge site, there is an obvious dependence of the Cu-S bond length on the different functionals, which cause it to vary from 2.25 Å to 3.34 Å. As in thiophene/Cu(111), PBE underestimates the interaction between the molecule and the surface, with a Cu-S bond length of 3.19 Å, while PBE-G06 overestimates this interaction and RPBE-G06 suppresses it again. The TS correction shows a slightly larger bond length than that of G06. For the vdW-DF series, vdW-DF2 gives the longest bond length (3.34Å), and optB86b the shortest (2.41Å), repeating the trend observed for the (111) surface. All of these methods show small or even negative tilting angles. In the IFS results, as the effective charge increases, the Cu-S bond length decreases from 3.38 Å (0.5$e$) to 2.85 Å (1.0$e$) and 2.63 Å (1.5$e$), while the tilt angle increases. These results indicate that the dominant contribution to the molecule-surface interaction for the Top-site configuration is due to Cu-S and Cu-C covalent bonds, whereas for the Bridge-site configuration the Cu-π interaction dominates.

It is somewhat surprising that all vdW-DF functionals favour the Top site. Recent scanning tunneling microscopy (STM) observations suggested that the Cu(100) surface is much more reactive than Cu(111) for phenyl-based molecules. One may speculate that during the XSW measurements [44], x-ray irradiation and thermal excitation may result in a dissociative attachment of thiophene on Cu(100), leaving a C$_4$ carbon chain/ring adsorbed on the surface and a single S atom located at a Bridge site. Such dissociative attachment has been observed for a similar molecule (TB-TTF, comprised of phenyl- and vinyl-groups and S atoms) on Cu(100) [54]. Further experimental and theoretical efforts are required, especially a direct low-temperature STM observation of thiophene adsorbed on Cu(100).

**Adsorption Energy.** We calculated the adsorption energies of thiophene adsorbed on all three surfaces using the different functionals (other than PBE-CH-1*e*) and compared these with the available experimental data, as summarized in Table 2. The experimental adsorption energy for thiophene on Cu(100) is -0.63 eV [43]. The adsorption energy for the other two surfaces is unavailable, but it can be estimated from other experiments. The adsorption energy of benzene on Cu(111) [51,52] is 20% lower than that on Cu(110) [53], as deduced from the temperature programmed desorption (TPD) temperatures of 225 K (111) and 280 K (110). The adsorption energy of a molecule on Cu(110) is usually very close to that of Cu(100), within a difference of 10%, as noted above. Thus one may infer that the adsorption energy of thiophene on Cu(111) is approximately 0.5 eV.

Table 2 shows that the PBE functional, as expected, underestimates the adsorption energies for all surfaces and sites, especially for the (111) surface. Adding either the G06 or the TS dispersion correction to PBE, however, considerably overestimates this energy, but leads to relatively accurate energies when applied to RPBE. This result can be ascribed to a better error cancellation between the overly attractive dispersion correction and a more repulsive exchange-correlation functional. The adsorption energies of the (111), (110) and (100) surfaces revealed by RPBE-G06 are -0.46 eV, -0.57 eV and -0.59(-0.52) eV (Table 2). By considering the correction to the zero-point energy, which is approximately 50 meV, the RPBE-TS values agree even better with the experiments. VdW-DF functionals also give very reasonable results. Although vdW-DF2 overestimates the Cu-S distance very strongly, the predicted adsorption energies are only some tens of meV smaller than the corresponding experimental values. The other three functionals slightly overestimate the adsorption energy, by values from 0.1 eV (optPBE-vdW) to 0.3 eV (optB86b-vdW). The sequence of energy values predicted by the four vdW-DF functionals is consistent with the Cu-S bond lengths.

The DFT-D methods, such as RPBE-G06, were found to be very robust for calculating adsorption energies. Table 3 shows adsorption energies calculated by RPBE-G06 for structures relaxed by all the methods considered in this work. The energy varies within 0.09 eV, 0.10 eV and 0.23 eV respectively for the (111), (110) and (100) surfaces, and thus is quite insensitive to the structural differences. The columns ``RPBE" in Table 3 list the DFT portion of the adsorption energy and columns ``G06" the contribution from the G06 dispersion correction. The DFT portion is always positive because RPBE is such a repulsive functional, while the attractive dispersion correction opposes this repulsion to yield a very reasonable adsorption energy. It is remarkable that the repulsion and attraction are enhanced or weakened simultaneously for most structures: the energy changes from RPBE and G06 are very similar over a wide range of adsorption distances, resulting in a rather ``stable" adsorption energy. In the case of thiophene, the molecule is predicted to bond to substrates more strongly than it physisorbs. There are some extreme cases where the molecule-substrate interaction is by physisorption, e.g. 6,13-pentacenequinone(P2O)/Ag(111) [55], in which the molecule-substrate distance optimized by PBE is 4.0 Å, but the RPBE-G06 value is 0.9Å shorter. Such a drastically shortened distance increases the adsorption energy only from 1.15 eV to 1.45 eV, indicating that the RPBE-G06 method gives reasonable adsorption energies for a wide range of molecule-substrate distances even for non-bonding interfaces.

**Electronic Density of States.** Experimentally measured electronic structures of metal-molecule interfaces are usually reproducible by standard DFT calculations, e.g. PTCDA/Ag(111) [6,29]. As discussed above, including dispersive contributions causes noticeable changes to the Cu-S bond length (*d*) and the molecular tilt angle (*a*), which can be expected to cause an appreciable variation of electronic structures. We have investigated this effect for both the Cu(111) and Cu(110) surfaces, because the largest difference of 0.8 Å among the predicted Cu-S bond lengths was found on the (111) surface and the smallest, ~0.1Å, on the (110) surface (Figure 2).

Figure 2(a) shows that structural differences scarcely modify the appearance of the DOS for the occupied states of the (111) interface. The largest shift of those states in the energy range within 2.0 eV below $E_{Fermi}$ is only 0.2 eV. However, the situation for the unoccupied states is more complicated. The LUMO-substrate hybridized state (*h*-LUMO) moves from approximately 2.0 eV (PBE/optPBE-vdW) to 1.5 eV (PBE-G06), although the Cu-S bond length is shortened by only 0.2 Å. The shorter Cu-S bond lengths correspond to lower-energy states. In addition, a band broadening is observable in the shaded area of Figure 2(a), which is enhanced as the bond length decreases, suggesting a strengthening thiophene-Cu(111) interaction. Although the structural difference between the two energetically degenerate (flat and tilted) configurations revealed by methods including RPBE-G06 and optPBE-vdW is quite significant (Table 1), Figure 2(b) indicates that both configurations have nearly identical DOSs. They differ only in that the band broadening vanishes and the energy of state *h*-LUMO moves slightly higher in the flat configurations. The structural differences revealed by the different methods for thiophene/Cu(111) have significant effects only on states in the shaded area and on state *h*-LUMO. In the case of thiophene/Cu(110), however, much smaller changes to the bond length give rise to almost unchanged DOSs in the energy window from -2.0 eV to 0.5 eV (Figure 2(c)). The gap between the two pronounced peaks located between 1.0 and 1.5 eV grows by 0.29 eV from the PBE (second top red line) to the PBE-G06 (bottom black line) results, and a new state appears at around 1.2 eV in the PBE-G06 results. Thus the DOS is determined not only by interfacial bond lengths but also by the combinations of exchange-correlation functionals and dispersion corrections employed.

The DOS shows no appreciable charge transfer from the Cu surfaces to thiophene, a situation different from PTCDA and other polycyclic aromatic hydrocarbons, such as pentacene, on Cu, Ag and Au surfaces [6-8,31,38,50]. In these systems, the lowest unoccupied molecular orbital (LUMO) interacts with substrate states and a charge transfer occurs from the surface to the LUMO, causing a Fermi-Level pinning (the

LUMO-surface bonding state is located just below the Fermi Level [6]). In thiophene/Cu, the $h$-LUMO state is located well above the Fermi Level, at 1.9 eV and 1.2 eV respectively for the PBE results on Cu(111) and Cu(110). However, the S atom bonds to the Cu surface through covalent Cu-S bonding and the hybridized states originate from the molecular LUMO and Cu surface states.

The position of the $d$-band centre is a well-established property which determines the reactivity of transition-metal surfaces [56]. The closer is the $d$-band centre to the Fermi Level, the more reactive the metal surface is expected to be. Table 4 summarizes the computed values of $E_{d\text{-band}}$ - $E_{Fermi}$ for the (111), (110) and (100) surfaces with three representative functionals, i.e. PBE, RPBE-G06 and optPBE-vdW. The positions predicted by these functionals are nearly identical for each surface, which indicates again that the valence band is insensitive to the choice of exchange functional in thiophene/Cu interfaces. The adsorption energies on different surfaces (considering the Top site for (100)) are, as expected, completely consistent with the position of the $d$-band centre. The only exception is the PBE result, where the (110) surface has a lower $d$-band centre and larger adsorption energy compared with the (100) surface, a result explicable by the generally poor performance of PBE in predicting adsorption energies.

**Charge density:** when vdW effects are included, the DOSs of all the interfaces largely preserve their original features, which implies no changes to the covalent or vdW nature of thiophene-Cu bonding. Figure 3 shows the differential charge densities (DCD), defined as $\rho_{DCD} = \rho_{Thiophene/Cu} - \rho_{Thiophene} - \rho_{Cu}$, for PBE, PBE-G06 and optPBE-vdW (flat) calculations on the (111), (110) and (100) interfaces. Charge reductions (cold colors) were found near both Cu and S (C) atoms. These reduced charges accumulate in a volume between the S atom and the Cu atom beneath it (hot colors), indicating a typical Cu-S covalent bonding picture for both interfaces, regardless of the method used in the calculation. Even if the Cu-S bond length is as large as 3.28 Å, characteristic covalent bonding features remain observable in the flat configuration of the optPBE-vdW-relaxed (111) interface, shown in Figure 3(c). In addition to the Cu-S bonding, similar Cu-C covalent bonding is also illustrated for thiophene/Cu(110) in Figure 3(d)-(f). Although a vdW bonding picture for thiophene/Cu(110) was suggested by a vdW-DF calculation on the basis of adsorption energies [22], ``covalent-like" bonding can be found in the electronic structure of $C_{60}$/Au(111) deduced by the same method [27]. The situation in thiophene/Cu(100) is similar to the other two interfaces, even though the discrepancy between theory and experiment over the adsorption site remains. In our work, optPBE-vdW explicitly shows covalent bonding, which implies that the other two vdW-DF methods, optB88-vdW and optB86b-vdW, should also suggest covalent bonding, because the Cu-S bond lengths they predict are shorter than in optPBE-vdW.

We close the presentation of our results by discussing the effects of core-hole interactions on the charge distribution. The IFS implementation of these interactions succeeded in reproducing the adsorption structure of thiophene-Cu(111) and Figure 4 shows the screening charge induced by the core-hole. This is concentrated primarily around the C atoms, consistent with the fact that electron-core-holes are created on these atoms. The accumulated charges transfer to the LUMO and thus strengthen the Cu-S bonding, resulting in a shorter Cu-S bond length and a larger tilt angle. No significant differences are visible among the four configurations in Figure 4, and core-hole contributions do not resolve the puzzle of the adsorption site for thiophene/Cu(100).

**Discussion**

The thiophene-Cu interaction predicted by PBE is rather weak: the binding energy is underestimated significantly and the Cu-S bond length slightly overestimated due to the missing, vdW correlation and dispersion correction. The inclusion of the G06 dispersion correction increases the binding energy too much, and the more repulsive RPBE functional is required to balance this correction in both the adsorption energy and the molecule-substrate separation. Dispersion corrections shorten this separation significantly in thiophene/Cu(111) and thiophene/Cu(100) at the Bridge site, where the molecule-substrate interaction is relatively weak, whereas for adsorption via the relatively strong Cu-S and Cu-C covalent bonds, as at the Top site in Cu(110) and Cu(100), the separation changes only slightly. The coefficients $C_6$ and $R_0$ in the TS correction were derived from GGA charge densities, which offer a better transferability, and thus it is recommended to use these if applicable. Our results, however, show that RPBE-TS overestimates slightly both the molecule-substrate separation and the adsorption energy. It has recently been shown that the inclusion of a repulsive many-body screening effect within TS results in a smaller adsorption energy [26].

We found that the choice of a proper exchange functional is crucial to the performance of vdW-DF. Our results show that the predicted adsorption energy varies inversely with the predicted molecule-substrate distance. The sequence of calculated adsorption energies is optPBE-vdW < optB88-vdW < optB86b-vdW for these three functionals sharing the same correlation functional, consistent with their behavior for small molecules and simple solids [18-20,24,35]. The vdW-DF2 method gives the weakest bonding strength. Among the four vdW-DF schemes, optPBE-vdW optimizes the adsorption energy and optB88-vdW the molecule-substrate distance compared with experiment; by contrast the best performance is achieved by optB88 for interaction energies of the S22 set [19] and by optB86b for lattice constants [20]. The poor performance of vdW-DF2 is not surprising because it was optimized to simulate molecules [18]. Our results suggest that it is still necessary to find an even

``better-match'' exchange functional - it should be as repulsive as optPBE, with the distance range where Pauli repulsion operates being similar to optB88.

If thiophene bonds tightly to the Cu surface, for example through covalent bonds at the Top site of Cu(110) and Cu(100), the molecule-substrate bonding is so strong that the bond length should not be affected significantly when the molecule is partially charged under x-ray irradiation. It is thus safe to compare the measured XSW results with the ground-state DFT results in these interfaces. Otherwise, for weaker interactions like the Cu-π interaction in thiophene/Cu(111) and thiophene/Cu(100) at the Bridge site, a small amount of transferred charge could drastically alter the molecular configuration. This is the reason why the configuration on the (111) surface is more sensitive to the consideration of IFS than that on the (100) surface at the Top site. The agreement between theory and experiment achieved by considering IFS for thiophene/Cu(111) extends our understanding of the relevance of core-hole interactions. It does not indicate that the DFT-D and/or vdW-DF methods employed here are inaccurate, but simply that the energy range of XSW measurements exceeds the reach of these ground-state methods. IFS is an appropriate means of extending these and also constitutes a probable solution for similar discrepancies observed on certain noble metal (111) surfaces [29,31-34]. These statements are, we believe, extendable to similar organic molecule-metal interfaces. However, IFS does not resolve the discrepancy regarding the adsorption site of thiophene/Cu(100), which calls for further experiments including neutron scattering and scanning tunneling microscopy, and also for a higher-level theoretical analysis.

Turning to the electronic structure of thiophene/Cu interfaces, the standard PBE functional reveals a covalent bonding picture in all cases on the basis of the differential charge density. This statement holds even if the predicted bond lengths and adsorption energies are quite different, demonstrating that these quantities may not be the only clues determining the bonding picture. The inclusion of dispersive contributions does not change the covalent picture to a vdW one, but the energies of certain unoccupied states and the band broadening depend strongly on the method used. Further measurements of electronic structures are required to refine the ``best'' theoretical basis for reproducing the DOS. All of our results shed considerable light on the physical and technical ingredients for accurate modelling of organic molecule-metal interfaces and are invaluable in refining the exchange functionals for DFT-D and vdW-DF methods.

## Methods

**DFT calculations.** Calculations were performed using the general gradient approximation (GGA) for the exchange-correlation potential [37] , the projector augmented wave method [45,46], and a plane-wave basis set as implemented in the

Vienna *ab-initio* simulation package [47,48]. The energy cut-off for the plane-wave basis was set to 400 eV for all configurations examined. Five layers of Cu atoms, separated by a 20 Å vacuum region, were employed to model the Cu surfaces. Supercells of sizes p(3×3), p(2×3) and p(3×3) were adopted to investigate the adsorption of thiophene on Cu(100), Cu(110) and Cu(111) surfaces, respectively. We anchored the S atom of thiophene on the Cu substrates with respect to the underlying sites, i.e., the Top, Bridge and Hollow sites, and rotated the molecule by the allowed symmetry operations of the different Cu surfaces. As a result, seven, six and six initial configurations were considered respectively for Cu(100), Cu(110) and Cu(111). Molecules were placed on one side of the slab with a dipole correction applied. A 6×6×1 *k*-mesh was adopted to sample all two-dimensional (2D) surface Brillouin zones for both geometry optimization and total energy calculation, and the results verified with one calculation using an 8×8×1 mesh. In geometry optimizations, all atoms except those for the bottom two Cu layers were fully relaxed until the residual force per atom was less than 0.02 eV/Å.

In the light of recent developments in vdW-DF techniques, which produce a significant reduction in their computational cost to the GGA level, both DFT-D and vdW-DF methods were employed in the structural relaxation and adsorption energy calculations. Dispersion corrections from the methods of Grimme (G06) [11] and Tkatchenko-Scheffler (TS) [13] were combined with two different GGA functionals, PBE [37] and RPBE [49], denoted respectively as ``PBE-G06", ``PBE-TS", ``RPBE-G06" and ``RPBE-TS". Four combinations of exchange and non-local correlation functionals were considered at the vdW-DF level, namely optPBE-vdW [19], optB88-vdW [19], optB86b-vdW [20] and vdW-DF2 [18].

**Ionic final state approximation.** The experimentally available interfacial geometry of organic molecule-metal interfaces, with which theoretical calculations can be compared, was measured using x-rays [29,30,40-42,44]. A core hole is created when x-rays excite a core electron from an atom of thiophene. Nearby electrons are prone to transfer to the lowest unoccupied state of the excited molecule, screening the core hole. The transferred electrons begin to relax to lower states and eventually fill the excited core level, eliminating the core hole. It has been suggested from studies of PTCDA/Ag(111) and $CuPcF_{16}$/Ag(111) [29,31-34] that the electron transfer process is much faster than the relaxation process of the transferred electrons. Thus if excitation events happen continuously and their separation is shorter than the relaxation time of the transferred electrons, some of these electrons are not fully relaxed. This results in a dynamic electron accumulation around the molecule, effectively making it become negatively charged. The geometry of a molecule in this effectively charged state, previously denoted IFS [31], can be very different from its neutral ground state. An accurate description of such a dynamic process for these interfaces may require time-dependent

DFT, which is a highly computationally demanding technique. However, it has been demonstrated that the IFS approximation, a much cheaper method, gives a good reproduction of experimental results for PTCDA/Ag(111) and CuPcF$_{16}$/Ag(111) when used with a proper effective charge [29,31-34]. We thus considered the structural variations of thiophene adsorbed on Cu surfaces with the IFS effect included. The term ``PBE-CH-*ne*'' is used to indicate an IFS state with a net effective transfer of *n* electrons to the C atoms of thiophene, in which *n* may be fractional.

## Acknowledgements

We are grateful to Prof. B. Normand for proofreading the manuscript. This work was financially supported by the National Natural Science Foundation of China (NSFC, Grant Nos. 11004244 and 11274380), the Beijing Natural Science Foundation (BNSF, Grant No. 2112019), the Ministry of Science and Technology (MOST) of China (Grant No. 2012CB932704) and the Basic Research Funds of Renmin University of China (RUC), provided by the Central Government (Grant Nos. 12XNLJ03 and 14XNH060). W.J. was supported by the Program for New Century Excellent Talents in Universities. Calculations were performed at the PLHPC of RUC and the Shanghai Supercomputer Center.

## Author contributions

W.J. conceived the project. Z.X.H. performed all vdW-DF calculations. Z.X.H. performed calculations for electronic properties at the DFT and DFT-D level and H.P.L. did structural relaxations. Z.X.H. prepared all figures. Z.X.H. and W.J. wrote the manuscript and H.P.L. reviewed it.

## Additional information

**Competing financial interests:** the authors declare no competing financial interests.

| Table 1 \| Cu-S bond length and molecular tilting angle of interfacial configurations on (111), (110) and (100) surfaces relaxed by different methods. | | | | | | | | |
|---|---|---|---|---|---|---|---|---|
| Method | (111)-Top-Site | | (110)-Top-Site | | (100) | | | |
| | Cu-S(Å) | ∠Mol(°) | Cu-S(Å) | ∠Mol(°) | Cu-S(Å) | | ∠Mol(°) | |
| | | | | | Top | Bridge[a] | Top | Bridge[a] |
| PBE | 2.68 | 21.6 | 2.39 | 16.4 | 2.47 | 3.19 | 14.6 | -4.4 |
| PBE-G06 | 2.51 | 7.6 | 2.34 | 3.0 | 2.33 | 2.28 | -1.4 | 0.3 |
| PBE-TS | 2.93 | 2.9 | 2.36 | 8.1 | 2.33 | 2.51 | -0.4 | -3.0 |
| RPBE-G06 | 2.62(3.17) | 13.2(-2.3) | 2.41 | 11.8 | 2.47 | 3.07 | 9.1 | -5.2 |
| RPBE-TS | 3.29 | -0.9 | 2.43 | 13.1 | 2.56 | 3.12 | 9.4 | -4.1 |
| optPBE-vdW | 2.88(3.28) | 12.3(2.2) | 2.45 | 10.8 | 2.54 | 3.01 | 10.2 | -2.3 |
| optB88-vdW | 2.61(2.74) | 17.5(9.4) | 2.45 | 16.0 | 2.47 | 2.83 | 8.2 | -3.2 |
| optB86b-vdW | 2.53 | 12.5 | 2.36 | 8.4 | 2.33 | 2.41 | 0.6 | -1.0 |
| vdW-DF2 | 3.44 | -1.4 | 2.74 | 10.5 | 2.99 | 3.34 | 6.1 | -4.7 |
| PBE-CH-1e | 2.56 | 22 | 2.32 | 20.7 | 2.39 | 2.85 | 21.7 | 17.0 |
| Exp. | 2.62±0.03[40] 2.50±0.02[41] | 26±5[40] | - | - | 2.42±0.02[44] | | 0±5[44] | |

Theoretical values are reported for the most energetically favored adsorption site. For the (111) interface, numbers in parentheses represent the corresponding values of the flat configuration. Both probable sites, namely Top and Bridge, are shown for the (100) interface.

[a] Vertical distance

| Table 2 \| Adsorption energies of thiophene adsorbed on Cu(111), (110) and (100) surfaces. | | | | |
|---|---|---|---|---|
| Method | (111) | (110) | (100)-Top | (100)-Bridge |
| PBE | -0.08 | -0.30 | -0.21 | -0.12 |
| PBE-G06 | -0.88 | -1.10 | -1.15 | -1.00 |
| PBE-TS | -0.99 | -1.26 | -1.20 | -1.09 |
| RPBE-G06 | -0.46 | -0.57 | -0.59 | -0.52 |
| RPBE-TS | -0.66 | -0.78 | -0.70 | -0.65 |
| optPBE-vdW | -0.65 | -0.82 | -0.77 | -0.69 |
| optB88-vdW | -0.73 | -0.88 | -0.88 | -0.78 |
| optB86b-vdW | -0.78 | -1.16 | -1.02 | -0.88 |
| vdW-DF2 | -0.45 | -0.53 | -0.51 | -0.48 |
| Exp. | - | - | -0.63 [43] | |

The unit of energy is eV. Results are listed for the top site on (111) and (110) surfaces and for both top and bridge sites on the (100) surface.

**Table 3 | Adsorption energy calculated by RPBE-G06 from structures optimized by different methods.**

| Str | (111)-Top-Site | | | (110)-Top-Site | | | (100)-Top-Site | | | (100)-Bridge-Site | | |
|---|---|---|---|---|---|---|---|---|---|---|---|---|
| | RPBE | G06 | RPBE-G06 | RPBE | G06 | RPBE-G06 | RPBE | G06 | RPBE-G06 | RPBE | G06 | RPBE-G06 |
| PBE | 0.22 | -0.64 | -0.42 | 0.08 | -0.64 | -0.56 | 0.21 | -0.79 | -0.58 | 0.21 | -0.74 | -0.52 |
| PBE-G06 | 0.65 | -1.03 | -0.38 | 0.44 | -0.96 | -0.52 | 0.76 | -1.16 | -0.40 | 0.77 | -1.10 | -0.33 |
| PBE-TS | 0.43 | -0.88 | -0.45 | 0.24 | -0.82 | -0.58 | 0.74 | -1.14 | -0.40 | 0.66 | -1.04 | -0.38 |
| RPBE-G06 | 0.32 | -0.78 | -0.46 | 0.14 | -0.71 | -0.57 | 0.27 | -0.86 | -0.59 | 0.30 | -0.82 | -0.52 |
| RPBE-TS | 0.26 | -0.70 | -0.44 | 0.11 | -0.69 | -0.58 | 0.31 | -0.87 | -0.56 | 0.31 | -0.82 | -0.51 |
| optPBE-vdW | 0.17 | -0.59 | -0.42 | 0.12 | -0.69 | -0.57 | 0.21 | -0.79 | -0.58 | 0.25 | -0.77 | -0.52 |
| optB88-vdW | 0.38 | -0.85 | -0.47 | 0.06 | -0.61 | -0.55 | 0.29 | -0.88 | -0.59 | 0.35 | -0.85 | -0.50 |
| optB86b-vdW | 0.02 | -0.47 | -0.45 | 0.45 | -0.96 | -0.51 | 0.61 | -1.08 | -0.47 | 0.67 | -1.04 | -0.37 |
| vdW-DF2 | 0.14 | -0.54 | -0.40 | 0.05 | -0.53 | -0.48 | 0.04 | -0.63 | -0.59 | 0.06 | -0.62 | -0.56 |
| Exp. | - | | | - | | | -0.63[43] | | | | | |

The unit of energy is in eV. Columns RPBE, G06 and RPBE-G06 refer to the DFT portion, dispersion correction and total value of the adsorption energy, respectively.

**Table 4 | Calculated energy differences between the d-band centre and Fermi level for (111), (110) and (100) surfaces.**

| | PBE | RPBE-G06 | optPBE-vdW |
|---|---|---|---|
| (111) | -2.43 | -2.46 | -2.44 |
| (110) | -2.40 | -2.31 | -2.32 |
| (100) | -2.31 | -2.29 | -2.34 |

The unit of energy is eV.

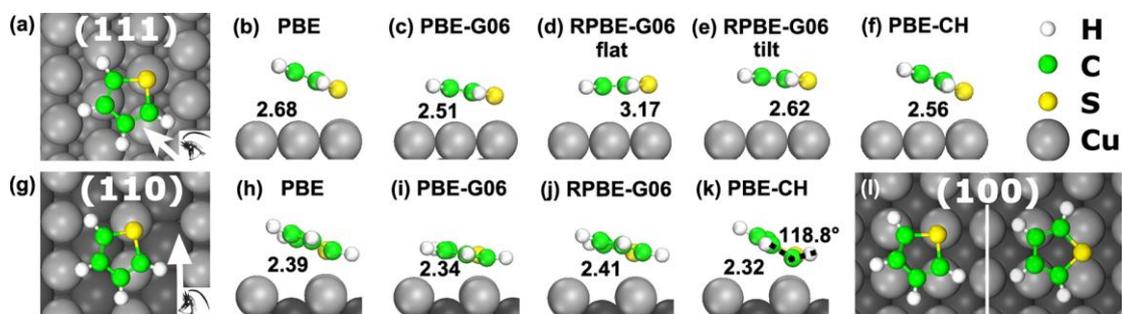

**Figure 1 | Top and side views of fully relaxed structures for thiophene molecules adsorbed on Cu surfaces.** For thiophene/Cu(111), all methods considered indicate that the molecule prefers to adsorb at the top site, as shown in the top view (a).

Side views show fully relaxed structures for thiophene/Cu(111) obtained using standard PBE (b), PBE-G06 (c), RPBE-G06 ((d) and (e)) and PBE with an electron-core- hole interaction (PBE-CH) (f). Values in the side views denote the Cu-S bond length in Å. For thiophene/Cu(110) (g) the top site is also preferred. Side views show the relaxed structures from PBE (h), PBE-G06 (i), RPBE-G06 (j) and PBE-CH (k). For thiophene/Cu(100), the two most probable adsorption sites, Top and Bridge, are shown in (l). All methods used reveal that the Top configuration is more stable than the Bridge by several tens of meV (Table 2).

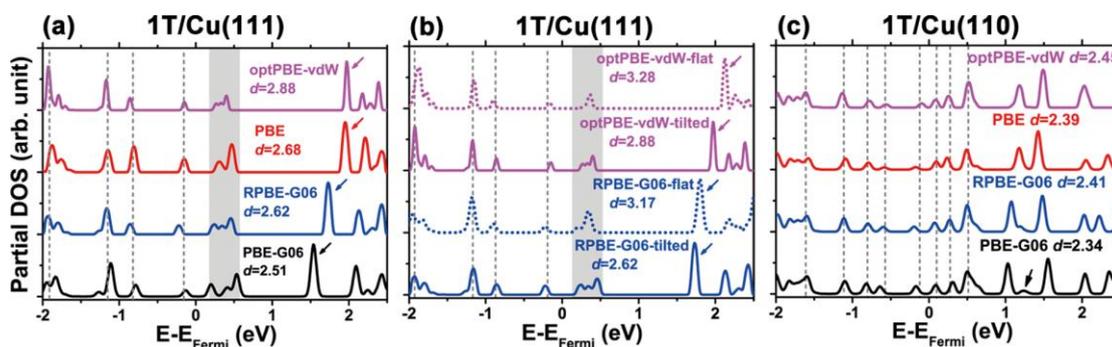

**Figure 2 | Partial local density of states (PLDOS).** PLDOSs of thiophene adsorbed on Cu(111) (a) and (b) and on Cu(110) (c) computed using PBE-G06 (bottom), RPBE-G06 (second bottom), PBE (second top) and optPBE-vdW (top). The comparison between the RPBE-G06 and optPBE-vdW results for the tilted (a) and flat configurations is made in (b). Variable ``$d$'' refers to the Cu-S distance, in units of Å. Arrows in (a) and (b) indicate the $h$-LUMO state and the black arrow in (c) indicates a state emerging with decreasing $d$.

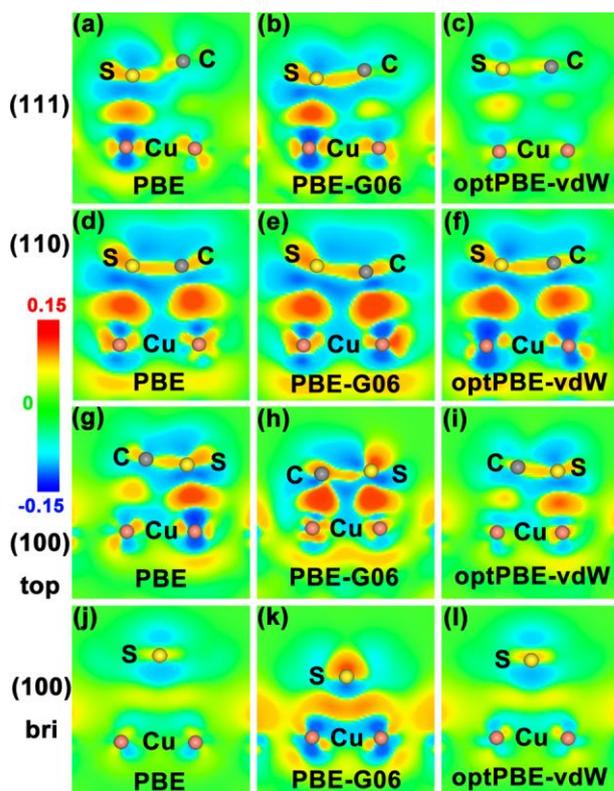

**Figure 3 | Differential charge density.** Results revealed by PBE ((a), (d), (g) and (j)), PBE-G06 ((b), (e), (h) and (k)) and optPBE-vdW((c), (f), (i) and (l)) for thiophene adsorbed on Cu(111) ((a)-(c)), Cu(110) ((d)-(f)) and Cu(100)((g)-(l)). Yellow, gray and brown balls represent S, C and Cu atoms, respectively. Slabs were cleaved along Cu-S or Cu-C bonds. Hot (e.g. red) and cold (e.g. blue) colors indicate respectively charge accumulation and reduction. . The scale of the colour map is not uniform, to enhance the contrast near 0. The unit is $e/Å^3$.

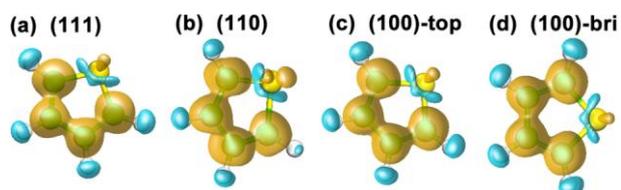

**Figure 4 | Charge density of core-hole-induced screening charge.** Charge redistribution after a core hole with an effective charge of $e$ is created. Isosurfaces of 0.03 $e/Å^3$ and -0.03 $e/Å^3$ show charge accumulation (orange) and reduction (blue).